\documentclass[11pt]{article}

\usepackage{epsfig}
\usepackage{cite}
\usepackage{amsmath,amsfonts,amssymb}
\usepackage{mathbbol}

\numberwithin{equation}{section}

\newcommand{\al}{\ensuremath{\alpha}}
\newcommand{\bt}{\ensuremath{\beta}}
\newcommand{\ga}{\ensuremath{\gamma}}
\newcommand{\Ga}{\ensuremath{\Gamma}}

\newcommand{\la}{\ensuremath{\lambda}}

\newcommand{\om}{\ensuremath{\omega}}
\newcommand{\Om}{\ensuremath{\Omega}}
\newcommand{\p}{\ensuremath{\psi}}

\newcommand{\N}{\ensuremath{{\cal N}}}

\newcommand{\ra}{\ensuremath{\rightarrow}}

\newcommand{\half}{\ensuremath{\frac{1}{2}}}

\newcommand{\quarter}{\ensuremath{\frac{1}{4}}}

\newcommand{\be}{\begin{equation}}
\newcommand{\ee}{\end{equation}}

\newcommand{\ba}{\begin{eqnarray}}
\newcommand{\ea}{\end{eqnarray}}

\newcommand{\ns}{\normalsize}

\newcommand{\gsim}{\raise.3ex\hbox{$>$\kern-.75em\lower1ex\hbox{$\sim$}}}
\newcommand{\lsim}{\raise.3ex\hbox{$<$\kern-.75em\lower1ex\hbox{$\sim$}}}

\newcommand{\nn}{\nonumber}
\newcommand{\wg}{\wedge}

\newcommand{\ir}[1]{\mathbf{#1}}
\newcommand{\irb}[1]{\overline{\ir{#1}}}
\newcommand{\irc}[1]{\ir{#1} + \irb{#1}}

\newcommand{\tc}[1]{\mathcal{W}_{#1}}

\newcommand{\Vol}{\mathcal{V}}

\newcommand{\Omn}{\Omega}
\newcommand{\Omb}{\overline{\Omega}}

\newcommand{\im}{\mathrm{Im}}
\newcommand{\re}{\mathrm{Re}}

\newcommand{\one}{\mathbb{1}}

\newcommand{\bg}[1]{\mathring{#1}}
\newcommand{\ev}[1]{\left< #1 \right>}

\newcommand{\href}[1]{\underline{#1}}

\begin{document}

\begin{titlepage}

\title{
   \hfill{\ns hep-th/0505177\\}
   \vskip 2cm
   {\Large\bf Effective action of (massive) IIA on manifolds with $SU(3)$ structure.}
\\[0.5cm]}
   \setcounter{footnote}{0}
\author{
{\ns \large 
  \setcounter{footnote}{3}
  Thomas House\footnote{email: thomash@sussex.ac.uk}, Eran Palti\footnote{email: e.palti@sussex.ac.uk} }
\\[0.5cm]
   {\it\ns Department of Physics and Astronomy, University of Sussex}
   \\
   {\ns Falmer, Brighton BN1 9QJ, UK} \\[0.2em] }

\date{}

\maketitle

\begin{abstract}\noindent

In this paper we consider compactifications of massive type IIA
supergravity on manifolds with $SU(3)$ structure.  We derive the
gravitino mass matrix of the effective four-dimensional ${\cal N}=2$
theory and show that vacuum expectation values of the scalar fields
naturally induce spontaneous partial supersymmetry breaking.  We go on
to derive the superpotential and the K\"ahler potential for the
resulting ${\cal N}=1$ theories. As an example we consider the $SU(3)$
structure manifold $SU(3)/U(1) \times U(1)$ and explicitly find $\N=1$
supersymmetric minima where all the moduli are stabilised at
non-trivial values without the use of non-perturbative effects.

\end{abstract}

\thispagestyle{empty}

\end{titlepage}

\section{Introduction}
\label{sec:introduction}

It has long been hoped that low-energy compactifications of string-
and M-theory will lead to phenomenological predictions which could be
tested experimentally.  A major obstacle to achieving this goal is the
presence of many light scalars, or `moduli', which are typically flat
directions of the four-dimensional theory. These arise as the massless
modes of the higher dimensional matter fields and as gauge-independent
variations of the metric on the compact space.  The precise values
taken by the moduli in the vacuum yield various parameters in the
four-dimensional model and therefore act as predictions coming from
string theory.  One of the most pressing concerns regarding string
theory compactifications, therefore, is the issue of moduli
stabilization.

One of the ways of inducing a non-trivial classical potential for the
low energy fields in the four-dimensional effective theory is through
the inclusion of non-vanishing field strengths for the ten-dimensional
fields with directions purely in the internal manifold. These are
referred to as fluxes and have been used extensively in the literature
for the purposes discussed above---see \cite{Michelson:1996pn,
Polchinski:1995sm, Louis:2002ny, Dall'Agata:2001zh} for some of the
earlier work.  Such fluxes back-react on the internal geometry and
will thus typically deform the internal space away from being Ricci
flat. In that case the Calabi-Yau manifolds used so often in
compactifications can cease to be a true solution of the theory. For
that reason compactifications on Calabi-Yau manifolds with fluxes are
restricted to the large volume limit where the fluxes are diluted and
their back-reaction may be ignored by treating them perturbatively.

Recently a growing body of literature has been looking at including
the back-reaction of the fluxes on the internal manifold and
considering manifolds which will be true solutions of the
theory. These manifolds will have non-vanishing torsion and so the
requirements on the compact space for preserving some supersymmetry in
the low energy theory is generalized from having special holonomy to
admitting a $G$-structure \cite{Gauntlett:2002sc}, which can be
classified in terms of its non-vanishing torsion classes.  

For M-theory most work is done on $G_2$-structure manifolds
\cite{Behrndt:2004bh, Behrndt:2004mx, House:2004pm, Lambert:2005sh},
although $SU(3)$-structure manifolds have also been considered
\cite{Lukas:2004ip, Dall'Agata:2003ir}. In the case of string theory
both $SU(3)$- and $SU(2)$-structure manifolds have been considered
\cite{Gurrieri:2004dt, Behrndt:2003ih, Behrndt:2004km, Behrndt:2004mj,
Lust:2004ig, Behrndt:2005bv}.  For a general review of structure
manifolds in string and M-theory see \cite{Gauntlett:2004hs} and
references therein.

From the point of view of phenomenology these manifolds have the
advantage that, although they are formally more general than
Calabi-Yau manifolds, they typically have a much simpler field
content. This can be thought of as the torsion placing restrictions on
the possible metric deformations of the manifold. They also have the
feature that, since they are not Ricci-flat, the four dimensional
background will not be Minkowski but (at least in the case where some
supersymmetry is preserved) anti-de Sitter.

This outcome is not desired for cosmological reasons and is the reason
that the Ricci-flat Calabi-Yau manifolds were originally more
attractive candidates. Recently, this reason has become less relevant
in the sense that when the inclusion of fluxes does produce a stable
vacuum in a Calabi-Yau compactification, that vacuum will typically be
anti-de Sitter anyway. It is therefore no more of a problem to start
from an anti-de Sitter cosmology in the first place and, like in the
Calabi-Yau case \cite{Kachru:2003aw}, hope that some non-perturbative
effects will lift this to a Minkowski or a de Sitter vacuum.

Another phenomenologically important feature of $SU(3)$-structure
manifolds is that known solutions on these manifolds to
ten-dimensional type IIA and IIB supergravities preserve ${\cal N}=1$
supersymmetry \cite{Behrndt:2003ih, Behrndt:2004km, Behrndt:2004mj,
Lust:2004ig, Behrndt:2005bv}, rather than ${\cal N}=2$ supersymmetry
which is problematic as a low-energy symmetry due \textit{inter alia}
to its lack of chiral representations.

In this paper we will consider compactifications of Romans' massive
type IIA supergravity on manifolds with $SU(3)$ structure.  An
important advantage of type IIA theory as opposed to type IIB is that
fluxes alone can generate non-trivial potentials for both complex
structure and K\"ahler moduli, although fully stabilising the moduli
has required non-perturbative effects such as instanton corrections
\cite{Kachru:2004jr}.  Recently, this has been overcome through the
use of orientifolds, where $\N = 1$ AdS solutions with all moduli
stabilised have been found \cite{Villadoro:2005cu, Derendinger:2004jn,
DeWolfe:2005uu}.

As yet, the covariant embedding of the massive IIA theory in the
M-theory `web of dualities' is not known, although it is believed to
encode information about the type-IIA string theory in D8-brane
backgrounds \cite{Bergshoeff:2001pv}.  The massive supergravity theory
is also considerably richer than the massless case, and we will not
concern ourselves with $\alpha '$ corrections or other `stringy'
effects that would require the full covariant embedding.

We will show that due to the torsion on the internal manifold there
are two types of fluxes that are associated with such
compactifications: the usual fluxes associated with non-perturbative
sources and fluxes originating from vevs of scalar fields. We will
then derive the effective low energy ${\cal N}=2$ theory by reducing
the gravitino mass terms to obtain the four-dimensional gravitino mass
matrix.  From this point we will restrict ourselves to the case where
the compact space is in a special class of half-flat manifolds shown
to be the most general manifolds compatible with the preservation of
$\N=1$ supersymmetry in four dimensions. We will show that for the
second type of fluxes the theory can exhibit spontaneous ${\cal N}=2
\ra {\cal N}=1$ partial supersymmetry breaking in the vacuum, and
construct the resulting $\N=1$ effective theory.

To study the vacua of the theory, we will consider a particular
compact space, and show that an $\N=1$ supersymmetric vacuum exists
where all the moduli are stabilised. This stabilisation does not
involve the introduction of any non-perturbative effects into the
superpotential or the use of orientifolds.

In section \ref{sec:su3structure} we summarise the relevant tools used
to classify $SU(3)$ structure manifolds and show how the structure can
be used to induce a metric on the internal manifold. In section
\ref{sec:iiareduction} we perform a Kaluza-Klein reduction of massive
type IIA supergravity to an $\N = 2$ effective four-dimensional theory
by deriving the gravitino mass matrix.  In section \ref{sec:n=1theory}
we will go on to derive the superpotential and K\"ahler potential of
the resulting ${\cal N}=1$ effective theory and show how the ${\cal
N}=2$ multiplets break to ${\cal N}=1$ superfields.  In section
\ref{sec:example} we will go through an explicit example of such a
compactification on the $SU(3)$-structure manifold $SU(3)/U(1)\times
U(1)$. We will derive the effective theory for compactification on
this coset and find an explicit supersymmetric minimum where all the
fields are stabilised at non-trivial vacuum expectation values.  We
summarise our conclusions in section \ref{sec:conclusion}.

\section{Manifolds of $SU(3)$ structure}
\label{sec:su3structure}

A six-dimensional manifold is said to have $SU(3)$ structure if it
admits a nowhere-vanishing two-form $J$ and three-form $\Omega$ (with
complex conjugate $\overline{\Omega}$) obeying conditions that we will
outline in this section. These forms are typically defined in terms of
bilinears of the Killing spinor $\eta_{+}$, which is a
nowhere-vanishing Weyl spinor of positive chirality with charge
conjugate $\eta_{-}$. Throughout this work, we use the spinor
conventions of \cite{Lust:2004ig} and so the spinors are normalised to
\be
\overline{\eta_+}\eta_+ = \overline{\eta_-}\eta_- = 1
\qquad \overline{\eta_+}\eta_- = \overline{\eta_-}\eta_+ = 0 \; .
\ee 
We can then write the structure forms as
\ba
J_{mn} & := & -i \overline{\eta_+} \ga_{mn} \eta_+ \nn \\
\Omega_{mnp} & := & \overline{\eta_-} \ga_{mnp} \eta \; ,
\label{su3def}
\ea
where $\ga^{m_1 \ldots m_n}$ denote anti-symmetric products of gamma
matrices. Note that although \eqref{su3def} makes use of the Killing
spinor, this spinor does not in fact contain the full information of
the structure forms, since the vielbein and hence the metric are
implicitly involved in their definition.

We shall now go on to consider the algebraic and differential
relations that the structure forms obey, as well as their relation to
the metric.

\subsection{Algebraic relations}
\label{sec:algebraicrelations}

Having defined the $SU(3)$ structure in terms of spinors and gamma
matrices, we can then use Fierz rearrangement formulae together with
commutation and anticommutation relations for gamma matrices to derive
algebraic relations such as
\ba
J_{m}^{\ p} J_{p}^{\ n} & = & - \delta_m^n \nn \\
J_{m}^{\ n} \Omega_{npq} & = & i \Omega_{mpq} \nn \\
(P_{+})_{m}^{\ n} \Omn_{npq} &=&  \Omn_{mpq} \nn \\
(P_{-})_{m}^{\ n} \Omn_{npq} &=&  0 \nn \\
\Omega \wg \overline{\Omega} & = & -\frac43 i J \wg J \wg J \nn \\
\Omega \wg J & = & 0  \nn \\
\star \Omn & = & -i \Om
\label{su3alg} \; ,
\ea
where $\star$ denotes the Hodge star and we have defined the usual
projectors
\be
(P_{\pm})_{m}^{\ n} := \frac12 ( \delta_{m}^n \mp i J_{m}^{\ n} ) \; .
\ee
We note that, in general, manifolds with $SU(3)$ structure are not
necessarily K\"ahler or even complex, despite the existence of a
globally defined almost complex structure $J$.  This means that the
usual distinction between holomorphic and anti-holomorphic indices
will no longer hold globally, however any local results obtained using
this distinction will still hold.

\subsection{Differential relations and torsion classes}
\label{sec:diffrelations}

$SU(3)$ structure also implies differential relations between the
structure forms $J$ and $\Omn$.  To derive these we first examine how
the deformation away from $SU(3)$ holonomy is parameterised.  The
contorsion $\kappa$ is defined, via the unique connection $\Ga_{T}$
that leaves the structure invariant, using the relation
\be
{\kappa_{mn}}^r := (\Ga_{T})_{[mn]}^{\ r} \; .
\ee
$\Ga_{T}$ is defined so that the derivatives given by it obey
\be
\nabla_{T} J = \nabla_{T} \Omega = 0 \ \ \ \ \ \ 
D_{T} \eta = 0 \; ,
\ee
where throughout we use $\nabla$ for a space-time covariant
derivative and $D$ for a spinor covariant derivative. Making use of
this relation, the Levi-Civita derivatives on the structure and spinor
are related to the contorsion via
\ba
(d J)_{mnp} & = & - 6 {\kappa_{[mn}}^r J_{p]r} \nn \\
(d \Omega)_{mnpq} & = &  12 {\kappa_{[mn}}^r \Omega_{pq]r} \nn \\
D_m \eta & = & \frac14 \kappa_{mnp} \ga^{np} \eta \label{contorsion} \; .
\ea
Note that since the contorsion is antisymmetric in its lowered indices,
it is still metric-compatible, so
\be
\nabla_{T} g_{mn} = \nabla g_{mn} = 0 \; .
\ee
The contorsion provides a way of classifying supersymmetric solutions
of supergravity theories by considering the structure group $G$ of the
manifold as a subgroup of $SO(N)$. Since in general the contorsion has
two antisymmetric indices and one other index, we have that
\ba
\kappa \in \Lambda^1 \otimes \Lambda^2
& \cong & \Lambda^1 \otimes so(n) \nn \\
& \cong & \Lambda^1 \otimes (g \oplus g^{\perp})\; ,
\ea
where $g$ is the Lie algebra on $G$ and $g^{\perp}$ is its complement
in $so(N)$. Since we know that the action of $g$ on the $G$-structure
must vanish by construction, we can decompose $\kappa$ according to
the irreducible representations of $G$ in $\Lambda^1 \otimes
g^{\perp}$. In the case of $SU(3) \subset SO(6)$, this gives
\ba
\kappa \in \Lambda^1 \otimes su(3)^{\perp}
& = & (\irc{3}) \otimes (\ir{1} + \irc{3}) \nn \\
& = & (\ir{1} + \ir{1}) + (\ir{8} + \ir{8}) + (\irc{6})
 + (\irc{3}) + (\irc{3}) \; .
\ea
We then associate each of these bracketed terms with a torsion class
$\tc{}$, which in concrete terms means that
\ba
dJ & = & -\frac32 \im (\tc{1} \overline{\Omega}) 
+ \tc{4} \wg J + \tc{3}  \nn \\
d\Omega & = & \tc{1} J \wg J + \tc{2} \wg J 
+ \overline{\tc{5}} \wg \Omega \; .  \label{torsionclasses}
\ea
The torsion classes can then be used to classify the structure
manifold.  In particular, for a manifold to be complex we need $\tc{1}
= \tc{2} = 0$, and where $\re(\tc{1}) = \re(\tc{2}) = \tc{4} = \tc{5}
= 0$ the manifold is \emph{half-flat}. The manifold that we shall go
on to consider will be half-flat but not complex. To see why such
spaces are not complex, we expand on our final comments in section
\ref{sec:algebraicrelations}, noting that for non-vanishing $\tc{1}$,
the relations in \eqref{torsionclasses} simply cannot be written in
holomorphic and anti-holomorphic coordinates.

A Calabi-Yau manifold thus has an alternative definition as a manifold
of $SU(3)$ structure with completely vanishing torsion
classes. Although in this sense, considering $SU(3)$ structure
manifolds with non-trivial torsion is more general than the Calabi-Yau
case, in fact the physics that we obtain from such manifolds will
often be simpler. For example, it was argued in \cite{Micu:2004tz}
that nearly-K\"ahler manifolds do not possess any complex structure
moduli, and this argument should also apply to the half-flat manifolds
that we will consider.

\subsection{Induced metric}
\label{sec:inducedmetric}

Having an $SU(3)$ structure on a manifold is a stronger condition then
having a metric.  In fact the forms $J$ and $\Omega$ induce a metric
on the space via the relation
\ba
g_{mn} & = & s^{-1/8} s_{mn} \mathrm{\ for} \nn \\
s_{mn} & = & - \frac{1}{64} ( \Omega_{mpq} \overline{\Omega}_{nrs} 
+ \Omega_{npq} \overline{\Omega}_{mrs} ) J_{tu} 
\hat{\epsilon}^{pqrstu} \label{indmet} \; ,
\ea
where $s$ is the determinant of $s_{mn}$. This form for the metric
allows us to express variations of the metric in terms of variations
of the $SU(3)$-structure
\ba
 \delta g_{mn} & = & 
  - \frac18 {(\delta \Omn)_{(m}}^{pq} \Omb_{n)pq}
  - \frac18 {(\delta \Omb)_{(m}}^{pq} \Omn_{n)pq}
  - (\delta J)_{t(m} {J^t}_{n)} \nn \\ & & 
  + \left[ \frac{1}{64} (\delta \Omn) \lrcorner \Omb 
  + \frac{1}{64} (\delta \Omb) \lrcorner \Omn 
  - \frac18 (\delta J) \lrcorner J \right] g_{mn}
 \label{indvar} \; .
\ea
In this form, calculation is rather difficult, however by using the
fact that $P_+ + P_- = \one$, we can obtain some of the
calculational convenience of holomorphic and anti-holomorphic
coordinates by acting on \eqref{indvar} with projectors to give
\ba
 {(P_+)_m}^p {(P_+)_n}^q \delta g_{pq} & = &
  - \frac18 \delta {\Omb_p}^{qr} {(P_+)_{(m}}^p \Om_{n)qr} \nn \\
 {(P_-)_m}^p {(P_-)_n}^q \delta g_{pq} & = &
  - \frac18 \delta {\Om_p}^{qr} {(P_-)_{(m}}^p \Omb_{n)qr} \nn \\
 \left[ {(P_+)_m}^p {(P_-)_n}^q + {(P_-)_m}^p {(P_+)_n}^q \right] \delta g_{pq} & = &
  - \frac18 \delta {\Om_p}^{qr} {(P_+)_{(m}}^p \Omb_{n)qr}
  - \frac18 \delta {\Omb_p}^{qr} {(P_-)_{(m}}^p \Om_{n)qr} 
  - \delta J_{p(m} {J_{n)}}^p \nn \\ & &
  + \left[ \frac{1}{64} (\delta \Omn) \lrcorner \Omb 
  + \frac{1}{64} (\delta \Omb) \lrcorner \Omn 
  - \frac18 (\delta J) \lrcorner J \right] g_{mn}
 \; . \label{projectvar}
\ea
Variations of the metric can, therefore, still be encoded in terms of
variations of $J$ and $\Om$, which we will refer to as K\"ahler and
complex structure deformations respectively.

\section{Reduction of the IIA action}
\label{sec:iiareduction}

In this section we will consider reducing the ten-dimensional action
for massive type IIA supergravity on a general manifold with $SU(3)$
structure. We will begin by summarising Romans' massive type IIA
supergravity. We will then show how to decompose the ten-dimensional
metric, Ricci scalar, dilaton, form fields and gravitino. Reducing the
terms that give gravitino mass terms will lead to an effective ${\cal
N}=2$ theory, which will be specified by the four dimensional
gravitino mass matrix.

\subsection{Action and field content}
\label{sec:fielddecomposition}

The action for massive type IIA supergravity, first outlined in
\cite{Romans:1985tz}, in the Einstein frame reads
\ba
S^{10}_{IIA} &=& 
\int\left( \nn
 \half \hat{R} \star 1  
 - \quarter d\hat{\phi} \wedge \star d\hat{\phi} 
 - \quarter e^{-\hat{\phi}} \hat{F}_3 \wedge \star \hat{F}_3
 - \quarter e^{\half \hat{\phi}} \hat{F}_4 \wedge \star \hat{F}_4 \right.\\ \nn
 &-& m^2 e^{\frac{3}{2}\hat{\phi}} \hat{B}_2 \wedge \star \hat{B}_2
 - m^2 e^{\frac{5}{2}\hat{\phi}} \star 1 \\ \nn
 &+& \left. \quarter d\hat{C}_3 \wedge d\hat{C}_3 \wedge \hat{B}_2
 + \frac{1}{6} m d\hat{C}_3 \wedge \hat{B}_2 \wedge \hat{B}_2 \wedge \hat{B}_2 
 + \frac{1}{20} m^2 \hat{B}_2 \wedge \hat{B}_2 \wedge \hat{B}_2 \wedge \hat{B}_2 \wedge \hat{B}_2
   \right) \\ \nn
 &+& \left. \int \sqrt{-\hat{g}} d^{10} X \right[
 - \hat{\overline{\Psi}}_M \Ga^{MNP} D_N \hat{\Psi}_P 
 - \half \hat{\overline{\la}} \Ga^M D_M \hat{\la}
 - \half (d \hat{\phi})_N \hat{\overline{\la}} \Ga^M \Ga^N \hat{\Psi}_M \\ \nn
 &-&\frac{1}{96} e^{\frac{1}{4}\hat{\phi}} (\hat{F}_4)_{PRST} \left( 
 \hat{\overline{\Psi}}^M \Ga_{[M}\Ga^{PRST}\Ga_{N]}\hat{\Psi}^N + 
 \frac{1}{2}\hat{\overline{\la}}\Ga^{M}\Ga^{PRST}\hat{\Psi}_M 
 + \frac{3}{8} \hat{\overline{\la}} \Ga^{PRST} \hat{\la} 
 \right) \\ \nn
 &+&\frac{1}{24} e^{-\frac{1}{2}\hat{\phi}} (\hat{F}_3)_{PRS} \left(
 \hat{\overline{\Psi}}^M \Ga_{[M}\Ga^{PRS}\Ga_{N]}\Ga_{11}\hat{\Psi}^N + 
 \hat{\overline{\la}}\Ga^{M}\Ga^{PRS}\Ga_{11}\hat{\Psi}_M \right) \\ \nn
 &+& \quarter m e^{\frac{3}{4}\hat{\phi}} \hat{B}_{PR} \left(
 \hat{\overline{\Psi}}^M \Ga_{[M}\Ga^{PR}\Ga_{N]}\Ga_{11}\hat{\Psi}^N 
 + \frac{3}{4}\hat{\overline{\la}}\Ga^{M}\Ga^{PR}\Ga_{11}\hat{\Psi}_M + 
 \frac{5}{8}\hat{\overline{\la}}\Ga^{PR}\Ga_{11}\hat{\la} \right) \\ 
 &-& \left. \half m e^{\frac{5}{4}\hat{\phi}}\hat{\overline{\Psi}}_M \Ga^{MN} \hat{\Psi}_N
 - \frac{5}{4} m e^{\frac{5}{4}\hat{\phi}}\hat{\overline{\la}} \Ga^{M} \hat{\Psi}_M
 + \frac{21}{16} m e^{\frac{5}{4}\hat{\phi}}\hat{\overline{\la}}\hat{\la} \right]
\label{10daction} \; .
\ea
This action is a generalisation of the type IIA supergravity that is
obtained from the low-energy limit of type IIA string theory, although
some care must be taken when taking the massless limit $m \rightarrow
0$ \cite{Romans:1985tz}. 

We now turn to notation and field content.  The indices $M,N \ldots$
run from 0 to 9, and the ten dimensional space-time coordinates are
$X^M$. In the Neveau-Schwarz-Neveau-Schwarz (NS-NS) sector the action
contains the bosonic fields $\hat{\phi}, \hat{B}_2, \hat{g}$, which
are the ten-dimensional dilaton, a massive two-form and the metric,
together with the fermionic fields $\hat{\Psi}, \hat{\la}$, which are
the gravitino and dilatino. The Ramond-Ramond (RR) sector contains the
three-form $\hat{C}_3$ and a one-form $\hat{A}^0$ which is eliminated
by a gauge transformation of $\hat{B}_2$ as in \cite{Romans:1985tz}.
The field strengths in the action are given by
\ba
\hat{F}_4 & := & d\hat{C}_3 + m \hat{B}_2 \wedge \hat{B}_2 \\
\hat{F}_3 & := & d\hat{B}_2 \; . \label{fieldstrenghts}
\ea
Note that, in contrast to the massless case, $\hat{F}_4$ will not in
general be closed, and that due to the equations of motion neither
field strength will in general be co-closed.

\subsection{Decomposing the metric}
\label{sec:decomposingmetric}
 
We now consider reducing the ten dimensional action on a manifold
endowed with $SU(3)$ structure. We split the ten dimensional
space-time coordinates as $(X^M)=(x^{\mu},y^{n})$ with external
indices $\mu,\nu \ldots = 0,1,2,3$ and internal indices $m,n \ldots =
4 \ldots 9$.  Reflecting the fact that we want the internal space to
be compact with compactification radii significantly smaller than any
length scales we wish to consider in four dimensions, we decompose the
ten-dimensional metric into a sum of four-dimensional and
six-dimensional metrics
\be
\hat{g}_{MN}(X) dX^M dX^N = \Delta(y) g_{\mu\nu}(x) dx^{\mu}dx^{\nu} 
+ g_{mn}(x,y) dy^m dy^n. \label{metricdecomp}
\ee
$\Delta(y)$ is a possible warp factor which will give the four
dimensional metric dependence on the internal coordinates.  In section
(\ref{sec:10dsolutions}) we will discuss the most general solution of
massive type IIA supergravity on manifolds with $SU(3)$ structure
\cite{Lust:2004ig} that preserves $\N = 1$ supersymmetry, where it was
shown that $\Delta(y)$ is in fact constant. We therefore consistently
set it to unity.  We note, however, that in the case where
supersymmetry is completely broken this warp factor may be non
vanishing.  \eqref{metricdecomp} also determines how the Dirac
matrices decompose and so we have
\be
\Ga^{\mu} := \ga^{\mu} \otimes \ga_7 \qquad
\Ga^m := \ga_5 \otimes \ga^m \; ,
\ee
where $\{\ga^{\mu}\}$, $\{\ga^m\}$ furnish representations of the
four- and six-dimensional Dirac matrices respectively.

\subsection{Ricci scalar reduction}
\label{sec:riccireduction}

Given the choice of metric ansatz above, the ten-dimensional Ricci
scalar can be written as
\be
 \hat{R} = R + R_6 - g^{mn} \nabla^2 g_{mn}
  - \frac14 g^{mn} g^{pq} 
  \left( \partial g_{mn} \cdot \partial g_{pq}
  - 3 \partial g_{mp} \cdot \partial g_{nq} \right) 
  \label{ricciexpand} \; ,
\ee
where $R, R_6$ are the four- and six-dimensional Ricci scalars
respectively. $\partial, \nabla$ are four-dimensional derivatives,
with $\cdot$ representing contraction over four-dimensional
indices. We shall reduce both the Einstein-Hilbert and dilaton kinetic
terms at the same time, which are given from \eqref{10daction} as
\be
S^{10}_{EH,D} = 
 \int d^{10} x \sqrt{-g} \left(
  \half \hat{R}
  - \quarter \partial_M \hat{\phi} \partial^M \hat{\phi}
 \right)
\label{s10ehd} \; .
\ee
Following the reduction of these terms, there are three field
redefinitions for the effective four-dimensional action that are
needed to put the kinetic terms in canonical form
\ba
 g_{\mu \nu} & \rightarrow & \Vol^{-1} g_{\mu \nu} 
  \label{weylrescaling} \\
 g_{mn} & \rightarrow & e^{- \hat{\phi} / 2} g_{mn} 
  \label{internalrescaling} \\
 \phi & := & \hat{\phi} - \half \mathrm{ln} \Vol 
  \label{redefs} \; ,
\ea
where $\phi$ is the four-dimensional dilaton. This gives a final form
for the four-dimensional action of
\be
S^4_{EH,D} =
 \int d^4 x \sqrt{-g} \left(
  \half R + \half e^{3 \phi / 2} \Vol^{-1/4} R_6 
  - \partial \phi \cdot \partial \phi
  + \frac{1}{8 \Vol} \int  d^6 x \sqrt{g} \partial g_{mn} \cdot \partial g^{mn}
 \right) \label{s4ehd} \; .
\ee
Our task is then to evaluate the internal integral in terms of the
$SU(3)$-structure forms, which is possible via the induced
metric, as discussed in section \ref{sec:inducedmetric}. This
allows us to write
\be
 \half \partial g_{mn} \cdot \partial g^{mn} =
  - \partial J_{mn} \cdot \partial J^{mn}
  + \frac18 \partial \Om_{mnp} \cdot \partial \Omb^{mnp} \; ,
\label{gdotg}
\ee
which we will use later in finding the K\"ahler potential.

\subsection{The Kaluza-Klein expansion forms}
\label{sec:decomposingstructure}

It was suggested in \cite{Gurrieri:2002wz}, and later developed in
\cite{D'Auria:2004tr, Tomasiello:2005bp} that a suitable basis for
Kaluza-Klein reduction on manifolds of $SU(3)$ structure is given by
two-forms $\om_i$, three-forms $\al_A, \bt^A$ and four-forms
$\widetilde{\om}^i$ obeying the algebraic relations
\ba
 \int \om_i \wg \widetilde{\om}^j = \delta_i^j & &
 \int \al_A \wg \bt^B = \delta_A^B \nn \\
 \int \al_A \wg \al_B & = & \int \bt^A \wg \bt^B = 0
\label{formsalg}
\ea
and the differential relations
\ba
 d \om_i & = & E_{iA}\beta^A - F_i^A\alpha_A \nn \\
 d \al^A & = & E_{iA} \widetilde{\om}^i \nn \\ \nn
 d \bt_A & = & F_i^A \widetilde{\om}^i \\
 d \widetilde{\om}^i & = & 0  
\; , \label{formsdiffgen}
\ea
where the matrices $E_{iA}$ and $F_i^A$ are constant.  In the limit
where $E_{iA},F_i^A\rightarrow 0$, we recover the usual set of
harmonic forms for a Calabi-Yau compactification:  $\{ \omega_i,
\widetilde{\omega}^j\}_{i,j = 1}^{h^{1,1}}$, $\al_0$, $\bt^0$, $\{
\al_a, \bt^b \}_{a,b = 1}^{h^{2,1}}$, where the $h^{p,q}$s are the
Hodge numbers of the manifold.  For the case where $E_{iA},F_i^A \neq
0$, however, it has been shown in \cite{Tomasiello:2005bp} that the
relevant forms do not carry topological information, and so there is
no metric-independent interpretation of the expansion forms.

Forms satisfying \eqref{formsalg} and \eqref{formsdiffgen} were shown
to be the correct basis for the case of half-flat manifolds with
Calabi-Yau mirror manifolds. It is natural to extend their use to
general half-flat manifolds, and it has been conjectured that such
forms could in fact be applied to general $SU(3)$-structure
compactifications \cite{Tomasiello:2005bp}. With this understood, we
shall proceed to make use of them whilst bearing in mind that other
bases for Kaluza-Klein reduction are not mathematically excluded.

\subsection{Decomposing the form fields and fluxes}
\label{sec:decomposingforms}

We decompose the ten-dimensional form fields in the following way:
\be
\hat{B}_2(X) = B(x) + \bg{B}(y) + b(x,y)
\ee
\be
\hat{C}_3(X) = C(x) + \bg{C}(y) + c(x,y) \; .
\ee  
Here $B$ and $C$ are external two and three-forms respectively.
$\bg{B}$ and $\bg{C}$ are internal two and three-forms with no
dependence on external co-ordinates. They give rise to NS-NS and RR
flux respectively.  $b$ and $c$ are two and three-forms that depend on
both the internal and external manifolds. Using the basis
(\ref{formsalg}) we can expand them as
\ba
 b(x,y) & = & b^i(x)\om_i(y) 
  \label{bexpansion} \\
 c(x,y) & = & \xi^A(x)\al_A(y) - \widetilde{\xi}_A(x)\beta^A(y) 
   + A^i(x) \wg \om_i(y)
  \; , \label{cexpansion}
\ea
where $A^i$ are space-time vectors.  Given our decomposition of the
form fields, the field strengths introduced in \eqref{fieldstrenghts}
can be written as
\ba
 \hat{F}_4 & := & d\hat{C}_3 + m \hat{B}_2 \wedge \hat{B}_2 \nn \\
   & = & d_4 (C + c) + d_6 (\bg{C} + c) +
   m (B + \bg{B} + b ) \wg (B + \bg{B} + b ) \nn \\
 \hat{F}_3 & := & d\hat{B}_2 \nn \\
   & = & d_4 (B + b ) + d_6 (\bg{B} + b ) \; ,
\ea
where $d_4$ and $d_6$ denote exterior derivatives on the external and
internal spaces respectively. We shall usually suppress these
subscripts. Of particular interest are the internal parts of the field
strengths (fluxes) which are given by
\ba
F_3 & := & d (\bg{B} + b) =: d B_2 \label{f3flux} \\ 
F_4 & := & d (\bg{C} + c) + m (\bg{B} + b) \wg (\bg{B} + b) \label{f4flux} \; .
\ea
In contrast to the usual situation for flux compactifications where
fluxes obtain their values entirely from the `background' field
strengths $\bg{B}$ and $\bg{C}$ (whose precise form is typically not
known) the fluxes here can receive contributions from the vacuum
expectation values (vevs) of the fields $b$ and $c$.%
\footnote{We note here that, as can be seen from \eqref{f3flux} and
\eqref{f4flux}, the splitting of the two types of contribution to the
flux is arbitrary. We could have defined $\bg{B}$ and $\bg{C}$ to
include the vevs of the scalars and then the scalars would have zero
vevs by definition. We have, however, chosen to keep the distinction
between the two types more apparent by considering $\bg{B}$ and
$\bg{C}$ as arising from sources other than the scalar vevs.}
This difference comes partly because the exterior derivative does not
automatically vanish on these fields and partly because the flux $F_4$
has a non-exact contribution from the second term in \eqref{f4flux}.
To distinguish between those fluxes that arise in the traditional way
and those that arise from vevs, we further define
\be
H_3 := d \bg{B} \qquad
G_4 := d \bg{C} + m \bg{B} \wg \bg{B} \; .
\ee
Now, to preserve Poincar\'{e} invariance of the four-dimensional
theory, all external components of the field strengths must be
proportional to the four-dimensional volume form.  This restricts us
to the only allowed external field strength of
\be
 (\hat{F}_4)_{\mu\nu\rho\sigma} 
  = f \epsilon_{\mu\nu\rho\sigma} 
 \; , \label{fdef}
\ee
where, due to its similarity with a similar parameter in the
eleven-dimensional case, we will call $f$ a Freud-Rubin parameter. The
Freud-Rubin parameter can be calculated in terms of the matter fields
by considering the dualisation of the external three-form $C(x)$.
Reducing the relevant terms in \eqref{10daction} gives the
four-dimensional action for $C(x)$
\be
S^{(4)}_C = \int_{X_4} \left[ -\quarter \Vol e^{\half \hat{\phi}} 
(dC + mB\wg B) \wg \star (dC + mB \wg B) + \half A dC \right] \; ,
\ee 
where
\be
 A := \int_Y \left[ 
  d\bg{C}\wg\bg{B} + b\wg d\bg{C} + dc\wg\bg{B} 
  + \frac{1}{3}m\bg{B}\wg\bg{B}\wg\bg{B} + m\bg{B}\wg\bg{B}\wg b
  + m\bg{B}\wg b \wg b + \frac{1}{3} m b \wg b \wg b 
 \right] \; . \label{Adef}
\ee
To dualise $C$ we follow the discussion in \cite{Micu:2003mp} and add
a Lagrange multiplier $\la$
\be
 S^{(4)}_C = \int_X \left[ -\quarter \Vol e^{\half \hat{\phi}} 
  (dC + mB\wg B) \wg \star (dC + mB \wg B) + \half A dC + \half \la dC \right]
 \; .
\ee 
Taking the equation of motion for $C$ and substituting in \eqref{fdef}
then gives
\be
 \star (dC + m B \wg B) 
  = \Vol^{-1} e^{-\half\hat{\phi}}\left( A + \la \right) 
  = - f
 \label{f} \; ,
\ee
which allows us to write $f$ in terms of the four-dimensional constant
$\la$ and the integral \eqref{Adef}.  We note here that we do not need
to perform a similar dualisation for $\hat{B}_2$, since although a
massless two-form would have been dual to a scalar, a massive two-form
will be dual to a massive vector, which we are not considering in our
analysis.

\subsection{The geometrical moduli}
\label{sec:N2multiplets}

We now turn to the fields arising from metric deformations.  From
general $\N = 2$ supergravity considerations \cite{deWit:1984pk,
Andrianopoli:1996vr}, we know that the complex structure deformations
$z^a$ span a special K\"ahler manifold ${\cal M}^{cs}$ with a unique
holomorphic three-form $\Omn^{cs}$, which has {\em periods} $Z^A$ and
$F_A ( Z^A )$, that are homogeneous functions of the $z^a$s.  The
K\"ahler potential is then given by the symplectic inner product
\be
 K^{cs} := - \ln i\left<\Omn^{cs} | \Omb^{cs} \right>   
         = -\ln i \left[ \bar{Z}^A F_A - Z^A \bar{F}_A \right] 
        =: -\ln ( ||\Omn^{cs}||^2 \Vol ) \; . \label{cskdef}
\ee
We can then use the forms in section \ref{sec:decomposingstructure} to
expand $\Omn^{cs}$ and re-write the K\"ahler potential as below
\ba
 \Omn^{cs} & = & Z^A \al_A - F_A \beta^A 
  \label{omegadecompose} \\
 K^{cs} & = & -\ln i\int \Omn^{cs} \wg \Omb^{cs} 
  \label{cskahlerpotential} \; .
\ea
We are now interested in relating $\Omn^{cs}$ to the holomorphic
three-form $\Omn$ in \eqref{su3def}. Writing
\be
\Omn^{cs} = \frac{1}{\sqrt{8}} ||\Omn^{cs}|| \Omn \label{omrelat} \; ,
\ee
we see that inserting \eqref{omrelat} into \eqref{cskahlerpotential}
and using \eqref{su3alg} we recover \eqref{cskdef}. As a check on this
process, we note that inserting the relation \eqref{omrelat} into
\eqref{projectvar} and going to a local patch where we can write
global holomorphic and anti-holomorphic coordinates, the usual
relations for metric variations are obtained
\be
 \delta g_{\al \overline{\bt}} = - i \delta J_{\al \overline{\bt}} 
  \; , \qquad
 \delta g_{\al \bt} = - \frac{1}{||\Omn^{cs}||^{2}}
  \left(  \delta  \Omb^{cs} \right){{}_{\al}}^{\ga \delta} 
  (\Om^{cs})_{\bt \ga \delta} \; .
\ee
The K\"ahler structure deformations $v^i$ arise in the usual way, after
we expand $J$ in the forms from section
\ref{sec:decomposingstructure}, which gives
\ba
 J & = & v^i\om_i \label{jexpansion} \\
 K & = & - \ln \frac{4}{3} J \wg J \wg J \label{tkahlerpotential} \; .
\ea
Inserting \eqref{bexpansion} into \eqref{10daction} and
\eqref{jexpansion} into \eqref{s4ehd} we see that the K\"ahler
structure deformations $v^i$ combine with the NS-NS scalars $b^i$ to
span a special K\"ahler manifold ${\cal M}^{SK}$ with K\"ahler
potential \eqref{tkahlerpotential}.

In summary, the geometrical moduli fields combine with the massless
modes of the matter fields to form ${\cal N}=2$ multiplets as shown in
Table \ref{N=2multiplets}.
\begin{table}
\center
 \begin{tabular}{||l|c||} \hline

 $g_{\mu\nu}, A^0$   & gravitational multiplet  \\ \hline
 $\xi^0, \widetilde{\xi}_0, \phi, B$  & tensor multiplet \\ \hline
 $b^i, v^i, A^i$ & vector multiplets \\ \hline
 $\xi^a, \widetilde{\xi}_a, z^a$  & hypermultiplets \\ \hline

\end{tabular}
\caption{Table showing the ${\cal N}=2$ multiplets in type IIA theory}
\label{N=2multiplets}
\end{table}   
The hypermultiplets span a quaternionic manifold ${\cal M}^{Q}$ with a
special K\"ahler submanifold ${\cal M}^{cs}$ and the vector multiplets
span the special K\"ahler manifold ${\cal M}^{K}$.

\subsection{Decomposing the gravitino}
\label{sec:decomposinggravitino}

Before we write down the mass matrix for the gravitini, we have to
choose an appropriate ansatz for the ten-dimensional gravitino.  As
discussed in section \ref{sec:su3structure} the internal manifold,
which has $SU(3)$ structure, supports a single globally defined,
positive-chirality Weyl spinor $\eta_+$ and its charge conjugate
$\eta_-$, which will have negative chirality. From standard arguments,
we expect terms involving other spinors on the internal space to lead
to four-dimensional masses at the Kaluza-Klein scale, and so they can be
ignored.  Given $\N = 2$ supersymmetry, we further expect the external
degrees of freedom for the gravitino to be given by a single Dirac
spinor which can be decomposed as two independent Weyl spinors. The
most general spinor ansatz for the ten dimensional gravitino that
involves these degrees of freedom is then
\be
\hat{\Psi}_M = \p_{M \al} \otimes ( a^{\al} \eta_+ + b^{\al} \eta_- )
+ \p_{M}^{\al} \otimes ( c_{\al} \eta_+ + d_{\al} \eta_- ) 
\label{gengravitino} \; ,
\ee
where the indices $\al, \bt$ are $SU(2)$ indices, which imply positive
chirality of a spinor when lowered and negative chirality when raised.
$a^{\al},b^{\al},c_{\al},d_{\al}$ are complex numbers.  $\p_{\mu 1,2}$
are thus four-dimensional gravitini with positive chirality and charge
conjugates $\p_{\mu}^{1,2}$, while $\p_{m 1,2}$ are four-dimensional
spin-$1/2$ fields with charge conjugates $\p_{m}^{1,2}$. Note that in
order not have cross terms between the gravitini and the spin-$1/2$
fields the gravitini need to be redefined with some combination of the
spin-$1/2$ fields. This does not affect the mass of the gravitini,
however, and so will not be considered here.

There are two physical constraints that we impose on the ansatz
\eqref{gengravitino} to restrict it. The first of these is that the
ten-dimensional gravitino should be Majorana. This gives the
conditions
\be
c_{1,2} = - (b^{1,2})^* \qquad d_{1,2} = - (a^{1,2})^* 
\; . \label{majoranagravitino}
\ee
The second constraint is that the gravitino ansatz should yield
canonical kinetic terms when reduced, which in this case look like
\be
S^{4}_{\mathrm{k.t.}} = - \int \sqrt{-g}d^{4}x \left(
\overline{\p^1_{\mu}} \ga^{\mu\rho\nu} D_\rho \p_{\nu 2}  
+ \overline{\p^2_{\mu}} \ga^{\mu\rho\nu} D_\rho \p_{\nu 2} 
 \right) + \mathrm{c.c.} \; ,
\ee
where c.c. stands for charge conjugate.  The kinetic term for the
ten-dimensional gravitino reads
\be 
S^{10}_{\mathrm{k.t.}} = \int \sqrt{-\hat{g}}d^{10}X
\left[\; - \hat{\overline{\Psi}}_M \Ga^{MNP} D_N \hat{\Psi}_P \; \right] 
\; . \label{10dkineticterm}
\ee
Substituting \eqref{gengravitino} into \eqref{10dkineticterm} and
performing the Weyl rescaling \eqref{weylrescaling} we get the
result that the four-dimensional gravitino kinetic terms will only
take the correct form when
\be
(a^{\al})^*(a^{\bt}) + (b^{\al})^*(b^{\bt}) 
= \frac12 \Vol^{-1/2} \delta^{\al \bt} 
\; . \label{canonicalkinetic}
\ee
Imposing \eqref{majoranagravitino} and \eqref{canonicalkinetic},
together with the absorption of a constant phase into one of
the spinor degrees of freedom, gives the following form for
the gravitino ansatz
\ba
\hat{\Psi}_M = \frac12 \Vol^{-1/4} & \bigg[ &
\p_{M 1} \otimes \left( \sqrt{1/2 + \varepsilon} \; \eta_+ 
+ \sqrt{1/2 - \varepsilon} \; e^{i \theta} \eta_- \right) 
\nn \\ & + & 
\p_{M 2} \otimes \left( \sqrt{1/2 - \varepsilon} \; \eta_+ 
- \sqrt{1/2 + \varepsilon} \; e^{i \theta} \eta_- \right) 
\bigg] + \mathrm{c.c.}
\ea
$\varepsilon$ can be chosen at convenience by making a further spinor
redefinition, while $\theta$ is a phase that is not fixed by physical
considerations and cannot be absorbed into a spinor redefinition.

Rather than leave these remaining parameters in, we note that upon
performing the reduction of terms that give a gravitino mass, it is
most convenient to choose $\varepsilon = 0$, while $\theta$ can be
eliminated by making the redefinitions below, which will not affect
the four-dimensional physics
\be
\Om \rightarrow e^{i \theta} \Om \qquad
M_{3/2} \rightarrow e^{i \theta} M_{3/2} \; ,
\ee
where $M_{3/2}$ is a gravitino mass. This gives us the working
ansatz for the gravitino 
\be
\hat{\Psi}_M = \frac{1}{2\sqrt2} \Vol^{-1/4} 
\left[ \p_{M 1} \otimes \left(\eta_+ + \eta_- \right)
+ \p_{M 2} \otimes \left(\eta_+ - \eta_- \right) \right]
+ \mathrm{c.c.} \label{gravitinoansatz}
\ee

\subsection{Gravitino mass matrix}
\label{sec:gravitinimass}

We are interested in the gravitino mass matrix of the ${\cal N}=2$
four-dimensional theory. The terms in the ten-dimensional action
(\ref{10daction}) which will contribute to the gravitino masses are
\ba
S^{10}_{\mathrm{mass}} = \left. \int \sqrt{-\hat{g}} d^{10} X \right[
 &-& \hat{\overline{\Psi}}_\mu \Ga^{\mu n \nu} D_n \hat{\Psi}_\nu \nn \\
 &-&\frac{1}{96} e^{\frac{1}{4}\hat{\phi}} (\hat{F}_4)_{prst}  
 \hat{\overline{\Psi}}^\mu \Ga_{[\mu}\Ga^{prst}\Ga_{\nu]}\hat{\Psi}^\nu \nn \\ 
 &-&\frac{1}{96} e^{\frac{1}{4}\hat{\phi}} (\hat{F}_4)_{\rho\sigma\delta\epsilon}  
 \hat{\overline{\Psi}}^\mu \Ga_{[\mu}\Ga^{\rho\sigma\delta\epsilon}\Ga_{\nu]}\hat{\Psi}^\nu \nn \\ 
 &+&\frac{1}{24} e^{-\frac{1}{2}\hat{\phi}} (\hat{F}_3)_{prs} 
 \hat{\overline{\Psi}}^\mu \Ga_{[\mu}\Ga^{prs}\Ga_{\nu]}\Ga_{11}\hat{\Psi}^\nu \nn \\  
  &+& \quarter m e^{\frac{3}{4}\hat{\phi}} \hat{B}_{pr} 
 \hat{\overline{\Psi}}^\mu \Ga_{[\mu}\Ga^{pr}\Ga_{\nu]}\Ga_{11}\hat{\Psi}^\nu \nn \\
 &-& \left. \half m e^{\frac{5}{4}\hat{\phi}}\hat{\overline{\Psi}}_\mu \Ga^{\mu\nu} \hat{\Psi}_\nu \right]
\; . \label{givegravitinomasses}
\ea
Using the ansatz (\ref{gravitinoansatz}), the definitions of $J$ and
$\Omega$ and the relations in section \ref{sec:su3structure} as well
as the discussion in section \ref{sec:decomposingforms} we can derive
the resulting four-dimensional masses.  After performing the
appropriate rescalings as in section \ref{sec:riccireduction} the mass
terms can be written as
\be
S^{4}_{\mathrm{mass}} =  \int \sqrt{-g} d^{4}x \left[ 
{S^{\alpha}}_{\beta} \overline{\p_{\mu \al}} \gamma^{\mu\nu} \p^{\beta}_{\nu} 
+ ({S_{\alpha}}^{\beta})^* \overline{\p^{\al}_{\mu}} \gamma^{\mu\nu} \p_{\nu \beta} 
\right] \; ,
\ee
where $\alpha,\beta=1,2$ label the gravitini. The mass matrix $S$ is
given by
\be
S = \left( 
\begin{array}{cc} 
M_1 & D \\ D & M_2 
\end{array} 
\right)
\ee
with terms defined as below
\ba
 M_1 & := & \frac{-i}{8} e^{2\phi} \Vol^{-\half} \Bigg[ 
  \la + \int_{Y} \left( d\bg{C} \wg \bg{B} + \frac{1}{3}m\bg{B} \wg \bg{B} \wg \bg{B} \right) 
  + \int_{Y} \left( dT + H_3 \right)\wg U  \nn \\
  & & + \int_{Y} \left( \frac{1}{3} m T \wg T \wg T + m \bg{B} \wg T \wg T + G_4 \wg T \right) \Bigg] \nn \\ 
 M_2 & := & - M_1|_{U \rightarrow \overline{U}} \nn \\
 D & := & \frac{-i}{8} e^{2\phi} \Vol^{-\half} 
  \int_Y (dT + H_3) \wg ( i \Vol^{-\half}e^{-\phi} \Omn^+ ) \nn \\
 T & := & b-iJ \nn \\
 U & := & c + i \Vol^{-\half}e^{-\phi}\Omn^-
  = c + i \sqrt{8} \Vol^{-\half} ||\Omn^{cs}||^{-1} e^{-\phi}\Omn^{cs-}
\; , \label{m1m2d}
\ea
where $\Omn^+$ and $\Omn^-$ are the real and imaginary parts of $\Omn$
respectively.  The four dimensional effective theory will be an ${\cal
N}=2$ gauged supergravity. Taking the general form for a gauged
supergravity found in \cite{Andrianopoli:1996vr} we see that using
their conventions the gravitino mass matrix is given by
\be
S_{\al \bt} = \frac{i}{2} P^x_{A} \sigma^x_{\al \bt} L^A
\label{genmassmatrix} \; ,
\ee
where the $P^x_{A}$ are prepotentials, $\sigma^x_{\al \bt}$ are Pauli
matrices and $L^A := e^{\half K^{cs}} Z^{A}$, where $K^{cs}, Z^{A}$
are defined in section \ref{sec:N2multiplets}.  Comparing
(\ref{genmassmatrix}) with (\ref{m1m2d}) we can completely determine
the K\"ahler potential of the vector multiplet sector and the
prepotentials of the hypermultiplet sector. We will not go on to do
this because in the next section we will see that quite generally this
theory will not preserve ${\cal N}=2$ supersymmetry in the vacuum and
we will instead have to consider specifying an ${\cal N}=1$ effective
theory.

\section{Breaking to $\N = 1$}
\label{sec:n=1theory}

In this section we will explore the implications of the form of the
gravitino mass matrix found in the previous section.  In order to do
this we will specialise to the case where the internal manifold is a
particular class of half-flat manifolds.  To motivate this choice we
will review the most general supergravity solution of massive type IIA
on manifolds with $SU(3)$ structure that preserves some supersymmetry
constructed in \cite{Lust:2004ig}.  We will then go on to show that
for that class of manifold the low energy theory will not preserve
${\cal N}=2$ supersymmetry in the vacuum and in fact will exhibit
spontaneous partial supersymmetry breaking to ${\cal N}=1$.  In
section \ref{sec:effectivetheory} we will derive the effective action
of the resulting ${\cal N}=1$ theory.

\subsection{Ten-dimensional massive IIA solutions}
\label{sec:10dsolutions}

In general, the reduction of type-II supergravities on spaces of
$SU(3)$ structure should yield an ${\cal N}=2$ supergravity.  There
are, however, solutions to (massive) IIA supergravity on manifolds of
$SU(3)$ structure that preserve supercharges consistent with ${\cal
N}=1$ supersymmetry in four dimensions. There were first considered in
\cite{Behrndt:2003ih, Behrndt:2004km, Behrndt:2004mj}, and later
generalised in \cite{Lust:2004ig}; we shall therefore refer to them as
BCLT (Behrndt-Cvetic-Lust-Tsimpis) solutions.  We present here a brief summary of the more general
solution in ten-dimensional language.  

The metric takes the form of \eqref{metricdecomp}, with $\Delta$
constant, while the fluxes and form fields for the solution take the
values
\ba
 m \hat{B}_2 & = & \frac{1}{18} f e^{- \hat{\phi} / 2} J + m \widetilde{B} \nn \\
 \hat{F}_3 & = & \frac45 m e^{7\hat{\phi} / 4} \Om^+ \nn \\
 \hat{F}_4 & = & f \star \one_4 + \frac35 m e^{\hat{\phi}} J \wg J
\label{lustfields} \; ,
\ea
where $f$ and $\hat{\phi}$ are constant.  $\widetilde{B}$ encodes the
non-singlet part of $\hat{B}_2$ and so obeys $\widetilde{B} \wg J \wg
J = 0$, but is otherwise quite general. A key feature of the solution
is that all torsion classes of the compact space vanish except for
\ba
\tc{1} & = & - i \frac49 f e^{\hat{\phi} / 4} \nn \\
\tc{2} & = & - 2 i m e^{3 \hat{\phi} / 4} \widetilde{B}
\label{lusttcs} \; .
\ea
Manifolds specified by the torsion classes \eqref{lusttcs} are
half-flat, and will play an important role in upcoming sections where
we will restrict the internal manifold to lie in this class. We note
here that we can always use this type of `internal' information from a
solution in constructing four-dimensional effective actions. 

It is informative to see how the fluxes arise in this solution.
Considering the torsion classes \eqref{lusttcs} and the relation
\eqref{formsdiffgen} and comparing the fluxes \eqref{f3flux} and
\eqref{f4flux} with \eqref{lustfields}, we see that the solution
corresponds precisely to the case where the fluxes arise purely from
the scalar vevs.  This will be an important observation later on when
we consider what types of fluxes break supersymmetry.
A further result of the solution that we shall make use of is that
\be
  M_{3/2} = \Delta \left( \frac{\al}{|\al |}\right)^{-2} 
 \left[ - \frac15 m e^{5 \hat{\phi} / 4} + \frac{i}{6} f e^{\hat{\phi} / 4} \right]
 \; , \label{lustuseful}
\ee
where $M_{3/2}$ is the value of the four-dimensional gravitino mass
for this solution and $\al$ is a constant related to the spinor phase
$\theta$ that we discussed in section \ref{sec:decomposinggravitino}
and can be consistently set to unity.

\subsection{Spontaneous partial supersymmetry breaking}
\label{sec:susybreaking}

We now want to consider the case where the $D$ terms in the mass
matrix vanish. From \eqref{m1m2d} and \eqref{lusttcs} we see that for
half flat manifolds $d \Omn^+ = 0$ and
so the $D$ terms indeed vanish. The mass matrix diagonalises under
this constraint and we see that there appears a mass gap $\Delta M^2$
between the two gravitini given by
\ba
 \Delta M^2 & = & |M_2|^2 - |M_1|^2 \nn \\
   & = & \frac{1}{32} e^{3 \phi} \Vol^{-1} \Bigg[
    \int_Y F_3 \wg \Om^- 
    \int_Y \left( \frac13 m J\wg J \wg J + F_4 \wg J \right)
 \nn \\ & & +
    \int_Y dJ \wg \Om^- 
    \int_Y \left( \frac16 f e^{\hat{\phi}/2} J\wg J \wg J
           +m B_2 \wg J \wg J \right) \Bigg] \; . \label{massgap}
\ea
It is interesting to consider how this mass gap depends on the fluxes.
In massless type IIA supergravity such a mass gap requires both RR and
NS-NS fluxes to be non-vanishing \cite{Curio:2000sc} (despite the
subtleties in doing so, is it possible to see this by taking the
limits $dJ, m\rightarrow 0$ in \eqref{massgap} above).  We see that
this is not the case here. Either type of flux by itself will generate
a mass gap due to a non-vanishing Freud-Rubin parameter
\footnote{The case where the Freud-Rubin parameter vanishes will not
be a proper supergravity solution and so we do not consider it here.}.
Hence, given general fluxes, the masses of the gravitini are
non-degenerate. This implies that we no longer have ${\cal N}=2$
supersymmetry. Indeed such a mass gap corresponds to partial
supersymmetry breaking with ${\cal N}=2 \rightarrow {\cal N}=1$ for a
physically massless lighter gravitino or full supersymmetry breaking
with ${\cal N}=2 \rightarrow {\cal N}=0$ for a physically massive
lighter gravitino

In a Minkowski background, physically massless particles simply have
zero mass. In anti-de Sitter (AdS) backgrounds, however, physically
massless particles can have non-zero masses
\cite{Breitenlohner:1982jf, Breitenlohner:1982bm, Gunara:2003td}.
This is the case here and so although the masses $M_1$ and $M_2$ in
\eqref{m1m2d} are non-zero for non vanishing fluxes one of them may
still be physically massless. As we saw in section
\ref{sec:10dsolutions} fluxes which arise from vevs can preserve
${\cal N}=1$ supersymmetry and therefore have a physically massless
gravitino. We can then check that one of our gravitini is indeed
physically massless by substituting the solution described in section
\ref{sec:10dsolutions} into our mass matrix \eqref{m1m2d} and checking
that one of the gravitini has a mass corresponding to the gravitino
mass found in the solution.

Putting the solution \eqref{lustfields} into the gravitino mass matrix
and taking care with the rescalings in section
\ref{sec:riccireduction}, we find firstly that $D = 0$. This means
that $\p_{1,2}$ are both mass eigenstates, with eigenvalues that obey
\ba
 M_1 & = & \frac15 m e^{5 \hat{\phi} / 4} - \frac{i}{6} f e^{\hat{\phi} / 4} \nn \\
 M_2 & = & - 3 M_1 \; .
\ea
Comparison with \eqref{lustuseful} gives that $| M_1 | = | M_{3/2} |$.
We therefore see that for the BCLT background, a mass gap opens up for
the two gravitini such that the $\p_\mu^1$ is physically massless and
$\p_\mu^2$ is physically massive.  With a slight abuse of terminology
we shall therefore refer to the lower mass gravitino as massless and
the higher mass one as massive.

For an inexhaustive list of literature discussing partial
supersymmetry breaking see \cite{ Cecotti:1984rk, Cecotti:1985sf, 
Ferrara:1995xi, Andrianopoli:2001gm, Curio:2000sc, Louis:2002vy,
Taylor:1999ii}.  Following their discussions we briefly summarise how
the matter sector of the theory is affected by the breaking.  In the
${\cal N}=2$ theory the fields were grouped into multiplets as
described in Table \ref{N=2multiplets}. Once supersymmetry is broken
these multiplets should split up into ${\cal N}=1$ multiplets. The
${\cal N}=2$ gravitational multiplet will need to split into a ${\cal
N}=1$ `massless' gravitational multiplet and a `massive'
spin-$\frac{3}{2}$ multiplet
\be
 \left( g_{\mu\nu},\p_1,\p_2,A^0 \right) \ra 
  \mathrm{massless}\;\left( g_{\mu\nu},\p_1\right) \;\;+\;\; 
  \mathrm{massive}\;\left(\p_2,A^0,A^1,\phi_1,\phi_2 \right) \; .
\ee  
Here $A^1$ is a vector field which has to come from one of the vector
multiplets and $\phi_1$ and $\phi_2$ are spin-$\half$ Fermions which
come from a hypermultiplet.  The $N_V$ ${\cal N}=2$ vector multiplets
break into $n_v$ massless ${\cal N}=1$ vector multiplets and $n_c$
massless chiral multiplets (with the other fields forming massive
multiplets) such that the scalar components of the chiral multiplets
span a K\"ahler manifold ${\cal M}^{KV} \subset {\cal M}^{SK}$. The
$N_H$ ${\cal N}=2$ hypermultiplets break into $n_h$ massless ${\cal
N}=1$ chiral multiplets and $N_H - n_h$ massive chiral multiplets with
$n_h \leq \half N_H$. The scalar components of the massless chiral
multiplets span a K\"ahler manifold ${\cal M}^{KH} \subset {\cal
M}^Q$.  With mass gaps appearing throughout the matter spectrum we can
consider working with an effective ${\cal N}=1$ theory by integrating
out the higher physical mass modes. For the case of scalars and
fermions this amounts to setting them to zero thereby truncating the
matter spectrum of the theory. It is not immediately clear from the
above considerations exactly which fields to truncate, however we will
return to this question in section \ref{sec:effectivetheory} when we
construct the ${\cal N}=1$ effective theory.
 
It is interesting to consider the case where $\bg{B} = 0 = \bg{C}$ and
the flux arises solely from the vevs of the scalar fields.  Then any
vacuum of the truncated ${\cal N}=1$ theory where the scalars have
non-vanishing vevs for which $\Delta M^2 \neq 0$ will indeed be a
valid vacuum of the full ${\cal N}=2$ theory. We will use this
observation to find such vacua in section \ref{sec:susyminima}.

\subsection{The ${\cal N}=1$ effective theory}
\label{sec:effectivetheory}

We are interested in constructing the effective ${\cal N}=1$ theory of
the physically massless modes. To do this we must explicitly determine
how the ${\cal N}=2$ multiplets in Table \ref{N=2multiplets} break
into ${\cal N}=1$ superfields and which of these superfields are
physically massive or massless. The form of \eqref{m1m2d} suggests
that $T$ and $U$ are the correct variables to expand in the chiral
superfields. To prove this is the case we will need to show that these
superfields span a K\"ahler manifold with a K\"ahler potential which
matches the one that will be derived from the gravitino mass.

We now turn to the calculation of the $\N = 1$ superpotential and
K\"ahler potential.  In the effective ${\cal N}=1$ theory the
remaining gravitino mass can be written as
\be
M_{3/2} = e^{\half K} W \; , \label{n1gravmass}
\ee
where $K$ is the K\"ahler potential and $W$ is the superpotential of the
theory. It is only this K\"ahler-invariant combination of $W$ and $K$ that
has any physical significance, although it is still  natural to decompose 
\eqref{n1gravmass} as
\ba
 e^{\half K} & = & \frac{e^{2\phi}}{\sqrt{8}\Vol^{\half}} 
   \label{kahlerpotential} \\
 W & = & \frac{-i}{\sqrt{8}} \Bigg[ \la 
   + \int_{Y} \left( d\bg{C} \wg \bg{B} + \frac{1}{3}m\bg{B} \wg \bg{B} \wg \bg{B} \right)
   \nn \\ & &
  + \int_{Y}\left( \frac{1}{3} m T \wg T \wg T + 
   m \bg{B} \wg T \wg T + G\wg T + ( dT + H ) \wg U \right)
 \Bigg] \label{superpotential} \; .
\ea
This gives a general form for the superpotential and K\"ahler
potential coming from the $\N =1$ effective action following
spontaneous breaking of the $\N =2$ theory for massive IIA on
manifolds of $SU(3)$ structure.  The theory will also have D-terms
corresponding to the off-diagonal elements of the $\N =2$ gravitno
mass matrix, $D$ in \eqref{m1m2d}, which vanish for half-flat
manifolds.  We will now express $W$ and $K$ in four-dimensional
language, assuming that we can expand in the forms of section
\ref{sec:decomposingstructure} so that
\be
 T = T^i \om_i, \qquad
 U = U^A \alpha_A - \widetilde{U}_{A} \beta^A 
 \; . \label{formsexpansion}
\ee  
We can then interpret $T^i, U^A,\widetilde{U}_{A}$ as the scalar
components of chiral superfields, of which the superpotential should
be a holomorphic function.  Substituting \eqref{formsexpansion} into
\eqref{superpotential}, we can write the superpotential as
\be 
W = \frac{-i}{\sqrt{8}} \left[ \lambda' + G_i T^i + B_{ij}T^i T^j + k_{ijk} T^i T^j T^k 
    + H_A U^A + \widetilde{H}^A \widetilde{U}_{A} 
    + ( F_i^A \widetilde{U}_{A} - E_{iA} U^A ) T^i \right] 
 \; , \label{superpot}
\ee
where $ \lambda',G_i,B_{ij}, k_{ijk},H_A,\widetilde{H}^A$ are
four-dimensional constants given by six-dimensional integrals
\be
 \begin{array}{rclrcl}
  \lambda' & = & \lambda + \int_Y ( d\bg{B} \wg \bg{C} 
   + \frac13 m \bg{B} \wg \bg{B} \wg \bg{B} ) &
  k_{ijk} & = & \frac13 m \int_Y \om_i \wg \om_j \wg \om_k \\
  B_{ij} & = & m \int_Y \bg{B} \wg \om_i \wg \om_j &
  G_i & = & \int_Y G \wg \om_i \\
  H_A & = & \int_Y H \wg \alpha_A &
  \widetilde{H}^A & = & \int_Y H \wg \beta^A \; .
 \end{array}
\label{constdefs}
\ee
As was discussed in section \ref{sec:susybreaking} turning on fluxes
$\bg{B},\bg{C} \neq 0$ will, in general, break supersymmetry
further. In the case where supersymmetry is completely broken it does
not make sense to talk about superpotentials and superfields.  If
these fluxes are small relative to the flux originating from the
scalar vevs, however, then they can be perturbatively included in the
superpotentials \eqref{superpotential} and \eqref{superpot}.  We
therefore display \eqref{superpot} as an indication of the class of
effective theories that can be obtained from the compactification of
massive IIA supergravity on spaces of $SU(3)$ structure. These may be
of use in, for example, studying $\half$-BPS states of such theories
as in \cite{House:2004hv}.

To be sure of retaining ${\cal N} = 1$ supersymmetry we will only
consider fluxes originating from scalar vevs from now on. In that case
the superpotential can be written as
\be
 W = \frac{-i}{\sqrt{8}}\left[ \la + \int_{Y} 
  \left( \frac{1}{3} m T \wg T \wg T + dT \wg U \right) \right] 
 \label{simplesuperpot} \; .
\ee
Having determined the superpotential of the effective theory we can
consider the K\"ahler potential.  To prove that
\eqref{kahlerpotential} is indeed the correct K\"ahler potential of
the truncated theory we need to explicitly perform the truncation and
show that the remaining fields form ${\cal N}=1$ superfields, $T^i,
U_A, \widetilde{U}_A$, with the corresponding metric. In the K\"ahler
moduli sector it was shown in section \ref{sec:decomposingforms} that
indeed the scalars $b^i$ and $v^i$ combine into $T^i=b^i - iv^i$ with
K\"ahler potential \eqref{tkahlerpotential}. In the hypermultiplet
sector we have $N_H$ hypermultiplets with $4N_H$ real scalar
components which are to be truncated to $n_h$ chiral multiplets with
$2n_h$ real components.  It seems that the correct superfields to form
are then
\ba
 U^A & = & \xi^A + i \sqrt{8} \Vol^{-\half}e^{-\phi} 
  \im \left( ||\Omn^{cs}||^{-1} Z^{A} \right) \\
 \widetilde{U}_A & = & \widetilde{\xi}_A + 
  i \sqrt{8} \Vol^{-\half} e^{-\phi} 
  \im \left( ||\Omn^{cs}||^{-1} F_A \right) \; .
\ea 
Indeed this form for the superfields has been proposed in
\cite{Nekrasov:2004js}, and also derived in \cite{Grimm:2004ua} for
the case where the partial supersymmetry breaking is induced through
an orientifold projection.  In our case, however, things are more
simple. The internal manifold is a half-flat manifold which has
torsion classes
\be
\re(\tc{1}) = \re(\tc{2}) = \tc{3} = \tc{4} = \tc{5} = 0 \; ,
\label{hftc}
\ee
so the general relations for the proposed Kaluza-Klein basis
\eqref{formsdiffgen} reduce to
\ba
d \om_i & = & E_{i} \beta_0 \nn \\
d \alpha_0 & = & E_{i} \widetilde{\om}^i \nn \\
d \widetilde{\om}^i & = & 0 = d \beta^A = d \alpha_{A\neq 0} 
\; , \label{formsdiff}
\ea
for $E_i := E_{0i}$. Applying \eqref{formsdiff} to
\eqref{torsionclasses} we arrive at
\ba
 dJ & = & E_{i} v^i \beta^0  
  = -\frac{3}{2} \im\left( \tc{1} \right) \re \left( \Omb \right)  
  \label{nocs1} \\
 d\Omn & = & Z^0 E_i \widetilde{\om}^i
  = i \im \left( \tc{1} \right) J \wg J 
  + i \im \left( \tc{2} \right) \wg J
 \label{nocs2} \; .
\ea
Equation \eqref{nocs1} is the motivation behind the statement that the
special class of half flat manifolds under consideration do not have
any complex structure deformations associated with them. This means
that we only have the tensor multiplet and so we only have one chiral
superfield left in the truncated theory. This superfield will contain
the dilaton $\phi$ and either $\xi^0$ or $\widetilde{\xi}_0$. To
decide which of the two is to be truncated we can refer to
\eqref{nocs2}.  We see that for our case, $ \re(\Omn) \propto \beta^0$
and $\im(\Omn) \propto \alpha_0$. And therefore since only the
imaginary part of $\Omn$ appears in the effective ${\cal N}=1$ theory
we should truncate the field associated with $\beta^0$, that is
$\widetilde{\xi}_0$.  Using the restrictions discussed above we can
write the remaining superfield as
\be
U_0 = \xi^0 + i e^{-\phi} \left( \frac{-4iZ^0}{F_0}\right)^\half \;.
\label{usuperfield}
\ee
Now inserting (\ref{cexpansion}) into (\ref{10daction}) we get the
kinetic term
\be
 S_{kin}^{U} =   \int \sqrt{-g} d^{4}x \left[ 
  - \left( \frac{F_0}{-4iZ^0} \right)e^{2\phi} \partial_\mu 
  \left(\xi^0 + i e^{-\phi} \left( \frac{-4iZ^0}{F_0}\right)^\half \right) 
  \partial^\mu \left( \xi^0 - i e^{-\phi} 
  \left( \frac{-4iZ^0}{F_0}\right)^\half \right)  \right]
  \; .
\ee
We see that taking the second derivatives,
\be
 -\partial_{U^0}\partial_{\bar{U}^0} \ln\left[ \frac{e^{4\phi}}{8\Vol} \right] 
 = \left( \frac{F_0}{-4iZ^0} \right)e^{2\phi}  \; ,
\ee
and so \eqref{kahlerpotential} is indeed the correct K\"ahler potential
and \eqref{usuperfield} is the correct superfield.

\section{An example: ${SU(3)}/{U(1)\times U(1)}$}
\label{sec:example}

Having derived in section \ref{sec:effectivetheory} the form of the
${\cal N}=1$ effective theory on a general manifold with torsion
classes (\ref{hftc}), in this section we will look at an explicit
example of such a manifold.  Denoting the internal manifold by ${\cal
Y}$ we will consider the coset space
\be
{\cal Y} = \frac{SU(3)}{U(1)\times U(1)}. 
\label{coset}
\ee
In section \ref{sec:cosetgeometry} we will derive explicit expressions
for $J$, $\Omn$ and the expansion forms on ${\cal Y}$. We will then
consider the effective theory and derive the superpotential and K\"ahler
potential. Finally we will find supersymmetric minima where all the
superfields have non-trivial expectation values.

\subsection{Geometry of the coset}
\label{sec:cosetgeometry} 

In general, a coset manifold ${\cal Y} := {\cal G}/{\cal H}$, where
${\cal H} \subset {\cal G}$, can be given a non-coordinate basis by
taking the generators of ${\cal G}$ and removing the generators of
${\cal H}$ in a way that is consistent with the embedding of ${\cal
H}$ in ${\cal G}$. We can then construct tensor products of this
basis, and it turns out that tensors on the coset are heavily
restricted by imposing that they remain invariant under the action of
any element of ${\cal G}$.  This restriction allows us to write the
most general ${\cal G}$-invariant tensors that can exist on the coset.

The particular case ${SU(3)}/{U(1)\times U(1)}$ has been considered in
\cite{Mueller-Hoissen:1987cq}, where the two $U(1)$ subgroups are
naturally identified with the diagonal Gell-Mann matrices. It was
shown that the most general ${\cal G}$-invariant two- and three-forms
can be written as
\ba
 A_{(2)} & = & \alpha e^{12} + \beta e^{34} + \gamma e ^{56} \nn \\
 A_{(3)} & = & \delta (e^{136} - e^{145} + e^{235} + e^{246})
  + \epsilon (e^{135} + e^{146} - e^{236} + e^{245}) 
\label{cosetforms} \; ,
\ea
where the $\{ e^m \}$ form a basis on the coset space, $\alpha \ldots
\epsilon$ are complex coefficients and we define $e^{m_1 \ldots m_p}
\equiv e^{m_1} \wg \ldots \wg e^{m_p}$.  Furthermore, by considering
the most general ${\cal G}$-invariant symmetric two-tensor on ${\cal
Y}$, we can define the metric on the coset space to be
\be
 g_{mn} e^m \otimes e^n := a ( e^1 \otimes e^1 + e^2 \otimes e^2 ) 
  + b ( e^3 \otimes e^3 + e^4 \otimes e^4 )
  + c ( e^5 \otimes e^5 + e^6 \otimes e^6 )
 \label{cosetmetric} \; ,
\ee
where $a,b,c$ are real. These three real parameters are the metric
moduli of the space ${\cal Y}$, and we would like to relate them to
the K\"ahler and Complex Structure forms.  Our first step in doing
this will be to construct specialisations of the two- and three-forms
in \eqref{cosetforms} that obey \eqref{su3alg}, and will therefore be
suitable for interpretation as the $SU(3)$-structure forms. Since some
of the conditions of \eqref{cosetforms} involve the metric,
constructing suitable forms also involves \eqref{cosetmetric}, and in
fact uniquely determines the K\"ahler and Complex Structure forms in
terms of $a,b,c$. A check on this procedure comes from
\eqref{indmet}. Imposing these constraints the $SU(3)$-structure forms
are given by
\ba
J & = & - a e^{12} + b e^{34} - c e ^{56} \nn \\
 \Omega & = & e^{i\varphi} \sqrt{abc} \left[
  \left( e^{135} + e^{146} - e^{236} + e^{245}\right) 
  - i \left( e^{136} - e^{145} + e^{235} + e^{246}\right) 
 \right] \; ,
\label{jandomega}
\ea
where $\varphi$ is an arbitrary phase which we can set to zero with no
loss of generality, a choice that corresponds to choosing the torsion
class conventions in \eqref{torsionclasses}. Now, since the basis on
${\cal Y}$ is just a subset of the generators of ${\cal G}$, their
derivatives will be given in terms of the structure constants for
${\cal G}$, and provided the division by ${\cal H}$ has been performed
adequately these derivatives should remain within ${\cal Y}$.  Taking
derivatives of the forms in \eqref{jandomega} thus gives---as a
specialisation of the result in \cite{Mueller-Hoissen:1987cq}
\ba
 d J & = & - (a + b + c) ( e^{135} + e^{146} - e^{236} + e^{245} ) \nn \\
 d \Omega & = & 4 i  \sqrt{abc} ( e^{1256} - e^{1234} - e^{3456} ) \; .
\label{cosettc}
\ea
Comparing \eqref{cosettc} with \eqref{torsionclasses}, we see that
${\cal Y}$ belongs to the special class of half-flat manifolds defined
in \eqref{hftc}.  Having found the appropriate forms and relations for
$J$ and $\Omn$ we can go on to look for a basis of expansion forms
that satisfy \eqref{formsalg} and \eqref{formsdiff}. A consistent set
of forms is given by
\be
\omega_1 = - e^{12},\ 
\omega_2 = e^{34},\ 
\omega_3 = - e^{56}
\ee
\be
\widetilde{\omega}^1 = - e^{3456},\ 
\widetilde{\omega}^2 = e^{1256},\
\widetilde{\omega}^3 = - e^{1234}
\ee
\be
\alpha_0 = - e^{136} + e^{145} - e^{235} - e^{246} 
\ee
\be
\beta^0 = - \frac{1}{4} \left( e^{135} + e^{146} - e^{236} + e^{245} \right) \; .
\ee
Note that we have made the choice $E_1 = E_2 = E_3 = 4$, however it
would have been possible to choose different values for these
parameters had we redefined the forms accordingly, and so this choice
is simply for convenience. We have also chosen the normalisation
convention $\int_{\cal Y} e^{123456} = 1$ so that the volume of ${\cal
Y}$ is given by
\be
\Vol = abc \;.
\ee
The structure forms $J$ and $\Omn$ can be written in terms of this basis as
\ba
 J & = & a \omega_1 + b \omega_2 + c \omega_3 \nn \\
 \Omn & = & \sqrt{abc} \left( i \alpha_0 - 4\beta^0 \right)
 \; . \label{jandomegadecompose}
\ea
It is also worth noting that the torsion classes can be evaluated
explicitly in this example, and are given by
\ba
 \tc{1} & = & \frac{2i}{3}
  \frac{a + b + c}{\sqrt{abc}} \nn \\
 \tc{2} & = & \frac{4i}{3}
  \frac{1}{\sqrt{abc}} \left[
  a(2a - b - c) e^{12} -
  b(2b - a - c) e^{34} +
  c(2c - a - b) e^{56} \right]
 \; . \label{su3tcs}
\ea
We have therefore been able to derive all the physically relevant
quantities in terms of the real metric parameters $a,b,c$. We can now
derive the effective theory arising from a compactification on the
space ${\cal Y}$.
 
\subsection{The effective theory}
\label{sec:susyminima}

In section \ref{sec:cosetgeometry} above we showed that the space
${\cal Y}$ has three moduli associated with K\"ahler structure
deformations.  By comparing \eqref{jexpansion} with
\eqref{jandomegadecompose}, we are able to relate them to the metric
parameters
\be
v^1 = a, \ v^2 = b, \ v^3 = c \; .
\ee
There were no geometric moduli associated with complex structure
deformations.  In the effective theory we therefore have three
superfields $T^1, T^2, T^3$ from the K\"ahler structure sector and the
superfield $U^0$ coming from the tensor multiplet. Using the
decomposition of $\Omn^{cs}$ in \eqref{omegadecompose}, together with
\eqref{omrelat} and \eqref{jandomegadecompose}, gives $F_0 = -4iZ^0$,
and so the superfields are
\ba
 T^i & = & b^i - i v^i \nn \\
 U^0 & = & \xi^0 + i e^{-\phi} \; .
\ea
Our knowledge of the coset space also allows us to evaluate the
superpotential \eqref{simplesuperpot} and the K\"ahler potential
\eqref{kahlerpotential}, which become
\ba
 W & = & \frac{-i}{\sqrt{8}} \left[ 
   \lambda + 2m T^1 T^2 T^3 - 4\left( T^1 + T^2 + T^3 \right) U^0 
  \right] \label{cosetsuperpot}\\
 K & = & -4\ln \left[-i\half\left( U^0 - \bar{U^0} \right)\right] 
   - \ln \left[-i
   \left( T^1 - \bar{T}^1\right)
   \left( T^2 - \bar{T}^2\right)
   \left( T^3 - \bar{T}^3\right)
  \right]
\label{cosetkahlerpot} \; .
\ea
We have now completely specified the ${\cal N}=1$ low energy effective
theory on the space ${\cal Y}$. It is then natural to ask whether this
theory has a stable vacuum. It is a well known result that
supersymmetric minima are stable vacua. We therefore look for such a
minimum by examining the F-term equations for the superpotential
\eqref{cosetsuperpot}, which read
\ba
 D_{T^1} W & = & 2mT^2 T^3 - 4U^0 - \frac{W}{T^1 - \bar{T}^1} = 0 \nn \\ 
 D_{T^2} W & = & 2mT^1 T^3 - 4U^0 - \frac{W}{T^2 - \bar{T}^2} = 0 \nn \\
 D_{T^3} W & = & 2mT^1 T^2 - 4U^0 - \frac{W}{T^3 - \bar{T}^3} = 0 \nn \\ 
 D_{U^0} W & = & -4 \left( T^1 + T^2 + T^3 \right) - \frac{4W}{U^0 - \bar{U^0}} = 0  
\label{feqs} \; ,
\ea
where the K\"ahler covariant derivative is given by $D_T := \partial_T
+ (\partial_T K)$.  A solution to these equations can be found by
setting $T^1=T^2=T^3 =: T$.  In this case the equations simplify to
the form
\ba
  U^0 & = & \frac{1}{24T\overline{T}} \left( -T ( \lambda + 2m\overline{T}^3 ) 
            + 3\overline{T} ( \lambda + 2mT^3 ) \right) \\ 
     0 & = &  - 6mT^2\overline{T}^3  
           - \lambda T\overline{T} -2m\overline{T}T^4 
      + 3\lambda T^2 - 2\lambda \overline{T}^2 -4mT^3\overline{T}^2 
      + 12mT\overline{T}^4  \label{tequation} \; .
\ea
A physically sensible solution to \eqref{tequation} should satisfy $m,
e^K, e^{-\phi} > 0$. Imposing these conditions gives a unique solution
with $\lambda > 0$ where the vacuum expectation values for the
superfield components are
\ba
 \ev{b^1} = \ev{b^2} = \ev{b^3} & = & 
   - \frac{5^{\frac23}}{20} \left( \frac{\lambda}{m} \right)^{\frac13}\nn \\
 \ev{v^1} = \ev{v^2} = \ev{v^3} & = & 
   \frac{\sqrt{3}5^{\frac16}}{4}\left( \frac{\lambda}{m} \right)^{\frac13} \nn \\
 \ev{\xi^0} & = & 
   - \frac{5^{\frac13}}{20}\left(m\lambda^2\right)^{\frac13} \nn \\
 \ev{e^{-\phi}} & = & 
   \frac{\sqrt{3}5^{\frac56}}{20}\left( m\lambda^2\right)^{\frac13}
\label{solution} \; .
\ea  
It is easily shown that these values for the scalars satisfy the BCLT
equations \eqref{lustfields}. The scalar potential is
\be
V = e^K \left[ K^{I\overline{J}} D_I W D_{\overline{J}} 
 \overline{W} - 3\left| W \right|^2 \right]
 \; , \label{scalarpotential}
\ee
where $I,J\ldots =0,1,2,3$ label the superfields and $K_{I\bar{J}} :=
\partial_I \partial_{\bar{J}} K$ has inverse $K^{I\bar{J}}$.
Substituting \eqref{solution}, \eqref{cosetkahlerpot} and
\eqref{cosetsuperpot} into \eqref{scalarpotential} we see that the
cosmological constant in the vacuum is given by
\ba
 \ev{V} & = & -3 e^K |W|^2 \nn \\
 & =: & \Lambda \simeq 
   \frac{- 29.0}{\left( m \lambda^5 \right)^{\frac13}} 
 \label{adsback} \; ,
\ea 
and so the solution has an anti-de Sitter background.  Having found a
stable vacuum of the effective ${\cal N}=1$ theory the discussion in
section \ref{sec:susybreaking} further implies that this is also a
stable vacuum of the full ${\cal N}=2$ theory.  The fact that it is a
supersymmetric anti-de Sitter vacuum means that it is stable even if
it is a saddle point \cite{Breitenlohner:1982jf,
Breitenlohner:1982bm}.

The moduli are therefore all stabilised without the use of any
non-perturbative effects like instantons and gaugino condensation, or
orientifold projections. To our knowledge this is the first example of
such a vacuum.  Because the stable vacuum arises from vevs of the
scalar fields there is no freedom in choosing the flux parameters. The
vacuum is in fact determined in terms of only two real parameters
$\la$ and $m$. This sits in contrast with the case of fluxes arising
from branes, where the only handle on the generation of flux
parameters comes from statistical `landscape'-type considerations.

\begin{figure}
\center
\epsfig{file=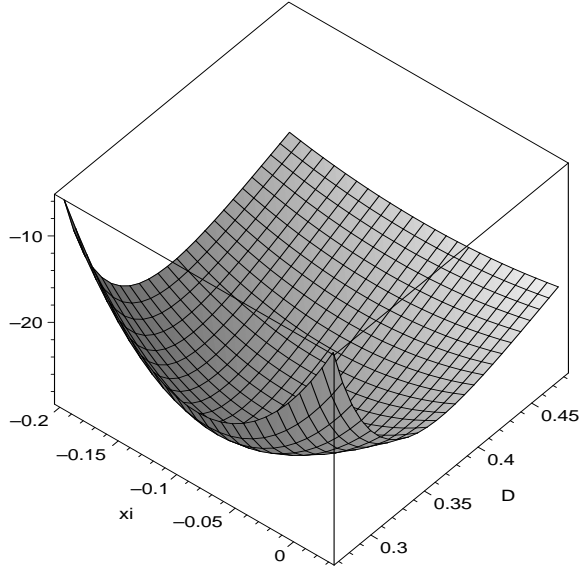,width=7.5cm,height=7.5cm}
\flushleft
\caption{Plot showing the scalar potential for the directions 
 $\xi^0$ and $e^{-\phi}$ (denoted as $D$). }
\label{fig:dilatonpot}
\end{figure}

\begin{figure}
\center
\epsfig{file=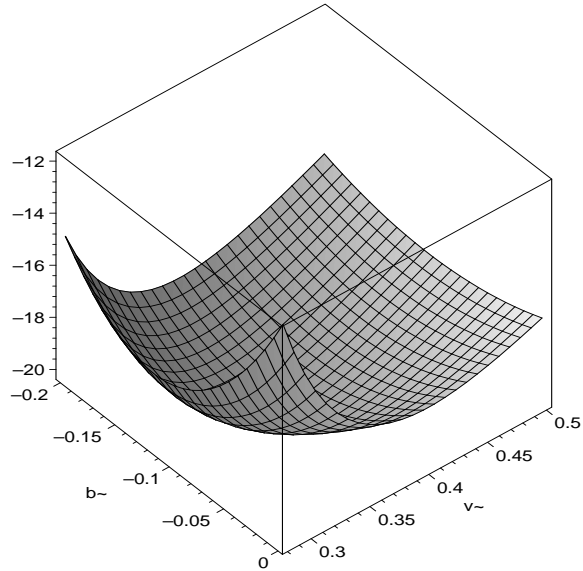,width=7.5cm,height=7.5cm}
\flushleft
\caption{Plot showing the scalar potential for the directions 
 $\widetilde{b}$ and $\widetilde{v}$.}
\label{fig:tpot}
\end{figure}

We may, however, eventually wish to consider uplifting the vacuum to a
Minkowski or a de Sitter vacuum through a mechanism similar to the one
used in the KKLT model \cite{Kachru:2003aw}.  Because such a possible
uplift will most probably involve non-perturbative effects and new
terms in the superpotential it may not leave the form of our solution
unchanged.  Nevertheless if an uplift leaves the solution unchanged
the question of whether it is a full minimum or a saddle becomes
important. We will therefore try to answer this question.  We can
construct a Hermitian block matrix from the second derivatives of the
potential with respect to the superfields evaluated at the solution
\ba
 H & := &  \left( \begin{array}{cc} V_{I\overline{J}} & V_{IJ} \\ 
   V_{\overline{I}\overline{J}} & V_{J\overline{I}} \end{array} \right) \\ 
   V_{I\overline{J}} &=& e^K K^{L\overline{M}}\partial_{L}\left( D_I W \right) 
   \partial_{\overline{M}} \left( D_{\overline{J}} \overline{W} \right) 
   - 2e^K K_{I\overline{J}}\left| W \right|^2 \\
 V_{IJ} &=& - \overline{W} e^K \partial_{I} \left( D_J W \right)  \; .
 \label{secondderivs}
\ea
Then for the solution to be a local minimum in all the directions
associated with the components of the superfields the matrix $H$ must
be positive definite. Inserting the solution \eqref{solution} into
\eqref{secondderivs} we find that out of the eight real eigenvalues
only six are positive. This means that there are two real directions
for which the potential is at a maximum. We can determine these
directions by looking at plots of the potential. Figure
\ref{fig:dilatonpot} shows the scalar potential for the two components
of the $U^0$ (axio-dilaton) superfield at constant $T^i$ with
$\lambda=m=1$.  We see that the potential forms a minimum with respect
these directions and so the maxima must be in directions associated
with the $T^i$ superfields. This raises the possibility that internal
spaces with different geometrical structure to ${\cal Y}$ may evade
this problem.  To illustrate this we may consider the potential with
the constraint $T^{1,2,3} =: \widetilde{T} =: \widetilde{b} -i
\widetilde{v}$ imposed. This would correspond to an internal space
with a single K\"ahler modulus, an example of which might be the coset
$G_2 / SU(3)$.  Figure \ref{fig:tpot} shows the scalar potential for
the directions associated with $T$ at constant $U^0$. We see that
again the potential forms a full minimum. Hence, although this is only
an indication of how things might go, it provides motivation for the
possibility of other spaces giving full minima and not saddles.

\section{Conclusions}
\label{sec:conclusion}

In this paper we have shown how the $\N = 2$ four-dimensional
effective action for (massive) IIA supergravity on manifolds of
$SU(3)$ structure can be constructed from the reduction of fermionic
terms. We then went on to show that it is possible to break $\N=2
\rightarrow \N=1$ spontaneously by having the scalar fields pick up
vevs. We derived the most general ${\cal N}=1$ effective theory that
can be obtained from such breaking. 

Using an example manifold we showed how it is possible to stabilise
all the fields in the vacuum without the use of any non-perturbative
effects or orientifold projections.  This is the first example we are
aware of where moduli are stabilised in this manner. The real
quantities $\la, m$ are the only free parameters in our solution,
which eliminates the need for statistical approaches to parameter
space such as the landscape.

The most obvious extension of this work is to look at different
explicit examples of half-flat manifolds. In particular other coset
manifolds and the Iwasawa manifold.  Another open question is whether
there are any systematic ways to study the moduli spaces of
$SU(3)$-structure manifolds that would determine whether our
assumptions about the basis for Kaluza-Klein reduction can be proved
for the general case.

Although our results depend on some specific features of the massive
IIA supergravity, it may be possible to obtain similar $\N = 2
\rightarrow \N = 1$ spontaneous breaking for other theories, for
example IIB and M-theory on manifolds of $SU(3)$ structure or type I
and heterotic string theories on manifolds of $SU(2)$ structure.
These manifolds offer several globally defined forms in terms of which
vev-derived fluxes could be written that might drive the super-Higgs
mechanism.

It would also be of interest, having stabilised the moduli, to study
the cosmology of the scalars as they roll towards the vacuum. There
are also the questions discussed earlier in this paper as to whether
the inclusion of non-perturbative effects could lift the vacuum to a
de Sitter background.

A further task for looking at phenomenology from these models would be
to look at getting a realistic particle content. This could be done
either through the use of intersecting branes or through the use of
spaces with both the conical singularities needed to obtain chiral
fermions, as in \cite{Acharya:2003ii}, and the $A-D-E$ singularities
needed to obtain gauge bosons.

\section*{Acknowledgments}

The authors would like to thank Andr\'e Lukas and Paul Saffin for
useful discussions and comments on this work.  We would also like to
thank Josef Karthauser for the use of his Mathematica{\textregistered}
code in verifying the results of section \ref{sec:cosetgeometry}.
Particular thanks must go to Andrei Micu for countless discussions,
suggestions and helpful observations.  Both authors are supported by
PPARC.

\appendix

\section{Conventions}

Throughout this paper we have used the space-time metric signature
$(-,+,+,...)$.  We define the $\epsilon$ symbol such that
$\hat{\epsilon}_{0123..} := +1$ with $\epsilon :=
\sqrt{|g|}\hat{\epsilon}$. The indices are raised and lowered with the
metric.  The components of a differential $p$-form $\om_p$ are defined
as
\be
 \omega_p = \frac{1}{p!} \omega_{\mu_1...\mu_p}
  dx^{\mu_1}\wg ... \wg dx^{\mu_p} \; .
\ee
The Hodge star operation $\star$ is defined such that
\be
 \omega_p \wg \star \omega_p = 
  \frac{\sqrt{|g|}}{p!}(\omega)_{\mu_1...\mu_{p}}
  (\omega)^{\mu_1...\mu_{p}}d^D x \; .
\ee
The contraction of a $p$-form and a $q \geq p$ form is given by
\be
 ( \om_p \lrcorner \Om_q )_{\mu_1 \ldots \mu_{q - p}} =
  (\om_p)^{\nu_1 \ldots \nu_{p}} 
  (\Om_q)_{\nu_1 \ldots \nu_{p} \mu_1 \ldots \mu_{q - p}} \; .
\ee
Dirac matrices anticommute to give
\be
\left\{ \Ga_M , \Ga_N \right\} = 2 g_{MN} \; .
\ee
Bilinears in spinors are constructed using the operation
$\overline{\p} = \p^{\dagger} \Ga^0$ for Minkowskian signatures and
$\overline{\p} = \p^{\dagger}$ for Euclidian signatures, where
${}^{\dagger}$ denotes Hermitian conjugation.


\end{document}